# Generation of optical frequency comb in a $\chi^{(2)}$ sheet micro optical parametric oscillator via cavity phase matching

Xinjie Lv[1*], Xin Ni[1*], Zhenda Xie[1+], Shu-Wei Huang[2], Baicheng Yao[3], Huaying Liu[1], Nicolò Sernicola[1,4], Gang Zhao[1], Zhenlin Wang[1], and Shi-Ning Zhu[1+]

[1] *National Laboratory of Solid State Microstructures, School of Electronic Science and Engineering, School of Physics, and College of Engineering and Applied Sciences, Nanjing University, Nanjing 210093, China*

[2] *Department of Electrical, Computer, and Energy Engineering, University of Colorado Boulder, Boulder, CO 80301, United States*

[3] *Key Laboratory of Optical Fiber Sensing and Communications (Education Ministry of China), University of Electronic Science and Technology of China, Chengdu 611731, China*

[4] *Institute for Optics, Information and Photonics, Friedrich-Alexander-Universität Erlangen-Nürnberg, Schloßplatz 4, Erlangen 91054, Germany*

[*] *These authors contributed equally to this work.*

+email: xiezhenda@nju.edu.cn
zhusn@nju.edu.cn

$\chi^{(3)}$ micro resonators have enabled compact and portable frequency comb generation, but require sophisticated dispersion control. Here we demonstrate an alternative approach using a $\chi^{(2)}$ sheet cavity, where the dispersion requirement is relaxed by cavity phase matching. 21.2 *THz* broadband comb generation is achieved with uniform line spacing of 133.0 *GHz*, despite a relatively large dispersion of 275.4 *fs²/mm* around 1064*nm*. With 22.6 % high slope efficiency and 14.9 *kW* peak power handling, this $\chi^{(2)}$ comb can be further stabilized for navigation, telecommunication, astronomy, and spectroscopy applications.

Coherent radiation with pristine frequency spacings, or optical frequency comb, is a ruler in optical frequency, which leads to the revolutionary precision metrology. It provides a direct link between microwave and optical frequencies [1] and can be utilized as an optical clockwork with unprecedented stability [2], and leads to broad applications in navigation [3], telecommunication [4,5], astronomy [6,7], and spectroscopy [8-13]. These applications benefit from miniaturized comb sources for portable devices with high beat note frequencies, which are not offered with the conventional mode-locked laser sources, but may be fulfilled by the recent development of micro optical parametric oscillators (μOPO). So far, most researches have been focused on the micro-ring μOPO with third order nonlinearity $\chi^{(3)}$ [14-20]. A high quality factor (Q) is the key for the enhancement of the relative weak third-order parametric processes in these micro-ring μOPOs to lower the parametric threshold. However, such high Q also makes the phase matching condition strict, because the effective nonlinear interaction length increases as the Q value scales up. Therefore, the dispersion needs to be perfectly engineered over a large comb span, which is normally a challenge for the μOPOs, and limit the comb central wavelength for a given material. On the other hand, parametric oscillation can also be generated using second order nonlinearity $\chi^{(2)}$ [21-28], which is much stronger than $\chi^{(3)}$ and reveals a new approach for μOPO comb. Our experimental and other earlier theoretical works show that the cavity phase matching (CPM) [29-35] can be realized in doubly-resonant $\chi^{(2)}$ Fabry-Perot micro-cavities [36], where the strict material phase matching is not necessary, as the phase mismatch is corrected by the total reflection on the cavity mirrors at each roundtrip. Although the effective nonlinear interaction length can still be greatly extended while Q increases, CPM condition only depends on the cavity length, and results in a large bandwidth.

Here such large CPM bandwidth plays a crucial role in this first $\chi^{(2)}$ comb demonstration with a monolithic sheet cavity. Two types of optical parametric oscillators (SOPOs) are fabricated with different cavity lengths. Broad comb spans are achieved from these SOPOs through a $\chi^{(2)}$ CPM optical parametric down conversion process, which exceed 21.2 *THz* and 14.3 *THz*, with comb line spacings of about 133.0 *GHz* and 514.0 *GHz,* respectively. Because of the strong second order nonlinearity, such sub-millimeter SOPO are capable of 22.6 % slope efficiency and 14.9 *kW* peak power. Despite the normal dispersion of 275.4 *fs²/mm*, comb lines are measured to be equidistant within the accuracy of a high-performance wavelength meter, which is less than the transform-limited single comb linewidth. These results reveal the line-to-line nonlinear interactions that automatically lock the line spacings, which are proved by our simulation results. This simulation also shows the potential for the future development of a fully stabilized comb in this $\chi^{(2)}$ SOPO platform.

As first proposed by Armstrong et al. in 1962 [29] and our earlier experimental demonstration [35,37,38], the sub-coherence-length Fabry-Perot nonlinear cavity can generate the second-order nonlinear process by cavity phase matching. The π phase shift gained from the reflection on the cavity mirrors compensates the phase mismatch of the light waves when they recirculates inside the cavity [33, 35], as the pump light is always in phase with the signal and idler when they recirculating inside the sheet cavity. Efficient frequency conversion can be achieved if the following first-order CPM condition is satisfied.

$$l_{Cav} \leq l_{Coh} = \pi / (k_p - k_s - k_i) \quad (1)$$

Where $l_{Cav}$ is the cavity length, $l_{Coh}$ is the coherence length of the three wave coupling, and $\vec{k}_p$, $\vec{k}_s$ and $\vec{k}_i$ are the wave vectors of pump, signal and idler, respectively. The effective nonlinear coefficient is given by:

$$d_{Eff} = d \left| \mathrm{sinc}(\frac{\pi l_{Cav}}{2 l_{Coh}}) \right| \quad (2)$$

In a doubly-resonant SOPO, both CPM and longitudinal mode matching conditions need to be satisfied for the resonance modes. Despite the large CPM bandwidth, the longitudinal mode matching results in strong frequency mode selectivity of the parametric beams. As shown in our previous work, tunable single-longitudinal-mode parametric oscillation can be achieved for SOPO with nondegenerate signal and idler modes either in polarization or in frequency. In this study, however, degenerate type-I CPM SOPO is chosen so that more than one pair of modes can be excited at the same time. As is shown in Fig. 1a, two types of SOPOs are fabricated from a monolithic crystal y-cut MgO doped lithium niobate crystal after fine polishing, with different thicknesses of 140 *μm* and 485.25 *μm*, respectively. The surface areas are both 5 *mm*×5 *mm* in the xz plane. The two end faces of these crystal sheets are both antireflection coated for 532 *nm* (reflectivity R<1%), and high-reflection coated in the 900 to 1150 *nm* range with R>99.8 % for the input surface and R=97.0 % for the output surface. The Q values are measured to be $6.6 \times 10^4$ or $1.9 \times 10^5$ for the 140 *μm* or 485.25 *μm* SOPOs, respectively. The type I CPM process is designed to be phase matched between the z-polarized pump at 532 *nm* and x-polarized parametric beams centered around 1064 *nm*. Based on the temperature dependent Sellmeier equation[39,41], simulations show that the above parametric down conversion process can be naturally phase matched at $114^\circ C$, and CPM can be used for extending the phase matching bandwidth, which is necessary for the comb generation. Fig. 1b shows the calculated coherence length in comparison to the cavity lengths (marked with red and blue lines) in the wavelength range from 900 *nm* to 1300 *nm*. It also includes the d$_{Eff}$ calculation for both SOPOs based on Eq. 2. According to Eq. 1, CPM bandwidths of 351 *nm* and 183.1 *nm* can be expected for the above two SOPOs, respectively.

In our experiment, the SOPOs are pumped by a single-longitudinal-mode frequency-doubled yttrium-aluminum-garnet laser (Powerlite, Continuum, Santa Clara, CA), with pulse duration of 10 *ns* and repetition rate of 10 *Hz*. A solid pinhole is put inside the laser cavity to purify the transverse mode. The output is focused onto another pinhole for further selecting the $TEM_{00}$ mode, and imaged onto the SOPO with a beam full width half maximum of 500 *μm*. The SOPO is embedded in a temperature-controllable oven with an accuracy better than 10 *mK*. The natural phase matching temperature for our SOPO is measured to be $111^\circ C$, which is in good agreement with the theoretical value. We scan the temperature around $111^\circ C$, and parametric oscillation output can be observed once the pump frequency is matched with the cavity modes for signal and idler. The single shot spectrum of the output is captured by a CCD-camera spectrometer (Acton 500, Princeton Instrument). As shown in Fig. 2a, b and c, the output beams show comb-like spectrum for both SOPOs, with spans exceeding 54 *nm* and 80 *nm*, or 14.3 *THz* and 21.2 *THz*, for the 140 *μm* and 485.25 *μm* SOPOs, respectively. The comb line spacings are measured to be 514.0 *GHz* and 133.0 *GHz*, which are in good agreements with the cavity lengths 140 $\mu m$ and 485.25 $\mu m$. Although the 140 $\mu m$ SOPO has a larger CPM bandwidth in theory, its higher threshold prevents us from achieving a broader comb span. It can be improved by further increasing the Q in future studies.

The thresholds for the comb generation are 860 $\mu J$ and 320 $\mu J$ for 140 $\mu m$ and 485.25 $\mu m$ cavities, with maximum output pulse energies of 42 $\mu J$ and 113 $\mu J$ at the pump energies of 1.03 $mJ$ and 830 $\mu J$, respectively. We measure the output pulse energy as a function of that of the pump for the 485.25 $\mu m$ SOPO. The result is shown in Fig. 2d. The maximum conversion and slope efficiency are calculated to be 13.6 % and 22.6 %, respectively. Under the condition of maximum output, the peak power of 14.9 $kW$ can be calculated considering the measured 5.5 $ns$ pulse width.

The uniformity of the comb spacing is the key feature of an optical frequency comb, and here it is studied in both experimental and theoretical approaches. Fig. 3a shows the schematic of the line-to-line measurement setup. The output parametric beam is collimated and directed to a high resolution double-pass grating filter, so that each single comb line can be filtered out individually. This grating filter comprises a high-efficiency ruled grating and a silver mirror, which reflects the first order diffraction beam back to the grating for the second diffraction. By tuning the mirror angle, we can couple any comb line into the fiber for the measurement. The optical frequency of each comb line is measured by a high-performance wavelength meter (WS-7, High finesse) with an absolute accuracy of 40 $MHz$. For simplicity, only the signal comb lines with higher frequency than degeneracy are measured, and we number them starting from #0 at the degenerate point. Our SOPO has a normal dispersion of 275.4 $fs^2/mm$ [38,39] at 1064 $nm$, however, the measured comb lines tend to be uniformly spaced, within the accuracy limit of the wavelength meter, as shown in Fig. 3b. The deviations of these comb line from the equal-spacing distribution are measured to be 7.6 $MHz$ in root-mean-square value. It agrees with the sub 10 $MHz$ relative accuracy of the wavelength meter, and is much smaller than the single comb linewidth, which is transform limited to about 200 MHz in such pulse-pumped case. This experiment result also shows good agreement with that of our simulation. We perform the simulation for such optical parametric effect in the SOPO, and only the second-order nonlinear optical effects are included in this simulation (For details, see supplementary information). In both experiment and simulation, the comb line spacing can deviate up to 1.049 *GHz* from the natural resonance of the SOPO cavity. More interestingly, some comb lines (#21-24th modes from the degeneracy) on the side of comb oscillate in the high-order transverse modes to follow the equal spacing, and stay within the cavity resonances. We have to couple them into a multi-mode fiber and sacrifice the accuracy to 160 *MHz*. This comb line pulling effect can only be interpreted as a result of the line-to-line nonlinear interaction. Such line-to-line interaction is well-known in the $\chi^{(3)}$ µOPOs, where two adjacent comb lines can seed into the next adjacent resonance modes via four wave mixing (FWM). Here this interaction may be interpreted as a result of cascaded $\chi^{(2)}$ processes [23-28], which is confirmed in our simulation with only second-order nonlinearity (For details, see supplementary information I and II). As shown in Fig. 3c, starting from comb line pair #+/-i ($\omega_{\pm i} = \omega_p/2 \pm i\Omega$, where $\Omega$ is the line spacing) and its closest adjacent line pair #+/-(i+1), the cavity enhanced sum frequency generation (SFG) process between $\omega_{-i}$ and $\omega_{i+1}$ can be cavity phase matched to generate $\omega_p + \Omega$. Then, difference frequency generation (DFG) happens between this SFG light and comb line #-(i+1), with output frequency of $\omega_i + 2\Omega$ into the #(i+2) resonance. This DFG process can seed into the optical parametric amplification (OPA) process with pump light for locking the #(i+2) line, and cascade over the whole comb span to lock the comb at one uniform spacing.

In the experiment, we use pulsed pump to demonstrate for first $\chi^{(2)}$ micro resonator comb generation in SOPO. The pulse profile limits the temporal uniformity, as well as the spectral resolution. With continuous-wave (CW) pump, the temporal and spectral noise may be further reduced, and it is confirmed in our simulation (For details, see supplementary information III). As shown in Fig. 4a , we scan the pump detuning $\upsilon$ at different pump levels to search for the lowest comb noise, where $\upsilon = \omega_p - 2\Omega_0$ and $\Omega_0$ is the central cavity resonance frequency. The comb intensity noise shows similar behavior as the detuning changes at different pump power, and the lowest noise can be achieve when $\upsilon = -193 MHz$ at 12 *kW* pump. We plot 266 repetitions of the temporal waveforms under above condition and the result is shown in Fig. 4b, and it can be found that these waveforms match each other very well. In fact, they are matched so well that we have to visually see them in the relative noise plot in Fig. 4c, and the relative noise is less than 0.13%.

The spectral behavior of the SOPO comb is also studied, and Fig. 4d is the simulated comb spectrum, with over 50nm span. Fig. 4e shows the simulation result of RF beat-note around the repetition frequency of 133.06 *GHz*, and the linewidth is fit to be 1.67 *MHz*. Such linewidth and the side band floor are limited by our simulation capability of 600ns waveform length, as proven by the comparison with the theoretically beat-note spectrum in the red line.

Our simulation shows a high CW oscillation threshold of 10 *kW* level for a bulk crystal SOPO used in the current experiment setup, which makes the CW pump impractical. However, the SOPO comb generation is possible with higher-Q and especially in waveguide devices, where the tight mode confinement further enhances the strong nonlinear interaction, thereby reducing the oscillation threshold, and the tailorable dispersion curve can result in a broader comb bandwidth. Our estimation shows that milliwatt-level CW-operation threshold can be achieved with a 500 µm long Lithium Niobate SOPO waveguide cavity [41-43] of $10^6$ Q.

In conclusion, we have demonstrated the first micro-resonator frequency comb generation based on second-order nonlinearity in a sheet optical parametric oscillator. The cavity phase matching plays a key role in mitigating the stringent phase matching condition imposed by a high-Q cavity and helps achieving a broader comb bandwidth up to 21.2 *THz*. The slope efficiency and peak output power exceed 22.6 % and 14.9 *kW*. Comb lines are measured to be equally spaced within the accuracy of the wavelength meter, 40 *MHz* (or 160 *MHz* in for comb lines #21-24), suggesting the possibility for a further active comb stabilization. The experimental results are also supported and confirmed by our computer simulations, which also reveal the capability of a broad-band low-noise comb generation with CW pump. Therefore, this new $\chi^{(2)}$ SOPO platform is a promising candidate for portable integrated optical frequency comb generation, which is required in navigation, communication, astronomy, and spectroscopy applications.

*Note added.* Recently we notice a related publication from S. Mosca et al. [44] in the latest journal of PRL after our resubmission. The authors independently demonstrate the $\chi^{(2)}$-based comb in a ring cavity configuration.

**Reference**

[1]  S. T. Cundiff and J. Ye，Rev. Mod. Phys. **75**, 325 (2003).

[2]  Th. Udem, R. Holzwarth, and T. W. Hänsch, Nature **416**, 233 (2002).

[3]  I. Coddington, W. C. Swann, L. Nenadovic, and N. R. Newbury, Nat. Photon. **3**, 351 (2009).

[4]  D. Hillerkuss *et al*., Nat. Photon. **5**, 364 (2011).

[5]  J. Pfeifle *et al*., Nat. Photon. **8**, 375 (2014).

[6]  C. H. Li, A. J. Benedick, P. Fendel, A. G. Glenday, F. X. Kärtner, D. F. Phillips, D. Sasselov, A. Szentgyorgyi, and R. L. Walsworth, Nature **452**, 610 (2008).

[7]  T. Wilken *et al*., Nature **485**, 611 (2012).

[8]  S. A. Diddams, L. Hollberg, and V. Mbele, Nature **445**, 627 (2007).

[9]  J. Mandon, G. Guelachvili, and N. Picque, Nat. Photon. **3**, 99 (2009).

[10]  B. Bernhardt, A. Ozawa, P. Jacquet, M. Jacquey, Y. Kobayashi, T. Udem, R. Holzwarth, G. Guelachvili, T. W. Hänsch, and N. Picqué, Nat. Photon. **4**, 55 (2010).

[11]  T. Ideguchi, S. Holzner, B. Bernhardt, G. Guelachvili, N. Picqué, and T. W. Hänsch, Nature **502**, 355 (2013).

[12]  A. Marian, M. C. Stowe, J. R. Lawall, D. Felinto, and J. Ye, Science **306**, 2063 (2004).

[13]  M. J. Thorpe, K. D. Moll, R. J. Jones, B. Safdi, and J. Ye, Science **311**, 1595 (2006).

[14]  P. Del'Haye, A. Schliesser,O. Arcizet, T. Wilken, R. Holzwarth, and T. J. Kippenberg, Nature **450**, 1214 (2007).

[15]  T. J. Kippenberg, R. Holzwarth, and S. A. Diddams, Science **332**, 555 (2011).

[16]  S. W. Huang, J. Yang, M. Yu, B.H. McGuyer, D. L. Kwong, T. Zelevinsky, and C. W. Wong,

Science Advances. **2**, e1501489 (2016).

[17]  S. Kim *et al.*, Nat. Commun. **8**, 372 (2018).

[18]  A. Dutt, C. Joshi, X. Ji, J. Cardenas, Y. Okawachi, K. Luke, A. L. Gaeta, and M. Lipson, Science Advances **4**, (2018).

[19]  D. T. Spencer *et al.*, Nature **557**, 81 (201 8).

[20]  A. B. Matsko, A. A. Savchenkov, D. Strekalov, V. S. Ilchenko, and L. Maleki, Phys. Rev. A **71**, 033804 (2005).

[21]  T. Beckmann, H. Linnenbank, H. Steigerwald, B. Sturman, D. Haertle, K. Buse, and I. Breunig, Phys. Rev. Lett. **106**, 143903 (2011).

[22]  J. U. Fürst, D. V. Strekalov, D. Elser, A. Aiello, U. L. Andersen, Ch. Marquardt, and G. Leuchs, Phys, Rev. Lett. **105**,263904 (2010).

[23]  I. Ricciardi, S. Mosca, M. Parisi, P. Maddaloni, L. Santamaria, P. D. Natale, and M. D. Rosa, Phys. Rev. A **91**, 063839 (2015).

[24]  S. Mosca, I. Ricciardi, M. Parisi, P. Maddaloni, L. Santamaria, P. D. Natale, and M. D. Rosa, Nanophotonics. **5**, 316 (2016).

[25]  T. Hansson, F. Leo, M. Erkintalo, S. Coen, I. Ricciardi, M. De Rosa, and S. Wabnitz, Phys. Rev. A **95**, 013805 (2017).

[26]  R. Ikuta, M. Asano, R. Tani, T. Yamamoto, and N. Imoto, Opt. Express. **26**, 15551 (2018).

[27]  X. Xue, F. Leo, Y. Xuan, J. A. Jaramillovillegas, P. H. Wang, D. E. Leaird, M. Erkintalo, M. Qi, and A. M. Weiner, Light. Sci. Appl. **6**, e16253 (2017).

[28]  Q. Ru, Z. E. Loparo, X. Zhang, S. Crystal, S. Vasu, P. G. Schunemann, and K. L. Vodopyanov, Opt. Lett. **42**, 4756 (2017).

[29]  J. A. Armstrong, N. Bloembergen, J. Ducuing, and P. S. Pershan, Phys. Rev. **127**, 1918 (1962).

[30]  E. Rosencher, B. Vinter, and V. Berger, J. Appl. Phys. **78**, 6042 (1995).

[31]  V. Berger, X. Marcadet, and J. Nagle, Pure Appl. Opt. **7**，319 (1998).

[32]  R. Haïdar, N. Forget, and E. Rosencher, IEEE J. Quantum Electron. **39**, 569 (2003).

[33]  Q. Clément, J. M. Melkonian, M. Raybaut, J. B. Dherbecourt, A. Godard, B. Boulanger, and M.

Lefebvre, J. Opt. Soc. Am. B. **32**, 52 (2015).

[34] M. Raybaut, J. B. Dherbecourt, J. M. Melkonian, A. Godard, M. Lefebvre, and E. Rosencher, Proc. SPIE. **8631**, 86311S (2013).

[35] Z. D. Xie, X. J. Lv, Y. H. Liu, W. Ling, Z. L. Wang, Y. X. Fan, and S. N. Zhu, Phys. Rev. Lett. **106**, 083901 (2011).

[36] A. Ciattoni, A. Marini, C. Rizza, and C. Conti, Light. Sci. Appl. **7**, 5 (2018)

[37] Y. H. Liu, Z. D. Xie, W. Ling, X. J. Lv, and S. N. Zhu, Opt. Lett. **36**, 3139 (2011).

[38] H. B. Lin, S. F. Li, Y. W. Sun, G. Zhao, X. P. Hu, X. J. Lv, and S. N. Zhu, Opt. Lett. **38**, 1957 (2013).

[39] D. E. Zelmon, D. L. Small, and D. Jundt, J. Opt. Soc. Am. B. **14**, 3319 (1997).

[40] A. L. Aleksandrovskiǐ, G. I. Ershova, G. Kh. Kitaeva, S. P. Kulik, I. I. Naumova, and V. V. Tarasenko, Sov. J. Quant. Elect. **21**, 225 (1991).

[41] E. Pomarico, B. Sanguinetti, N. Gisin, R. Thew, H. Zbinden, G. Schreiber, A. Thomas, and W. Sohler, New J. Phys. **11**, 113042 (2009).

[42] K. H. Luo, H. Herrmann, S. Krapick, B. Brecht, R. Ricken, V. Quiring, H. Suche, W. Sohler, and C. Silberhorn, New J. Phys. **17**, 073039 (2015).

[43] E. Pomarico, B. Sanguinetti, C. I. Osorio, H. Herrmann, and R. T. Thew, New J. Phys. **14**, 033008 (2012).

[44] S. Mosca, M. Parisi, I. Ricciardi, F. Leo, T. Hansson, M. Erkintalo, P. Maddaloni, P. De Natale, S. Wabnitz, and M. De Rosa, Phys. Rev. Lett. **121**, 093903(2018).

**Acknowledgements**

This work is supported in part by the National Young 1,000 Talent Plan; National Natural Science Foundation of China (No.91321312, No. 11621091，No. 11674169); Ministry of Science and Technology of the People's Republic of China (No. 2017YFA0303700); International Science and Technology Cooperation Program of China (ISTCP) (No. 2014DFT50230); and Key Research Program of Jiangsu Province (No. BE2015003-2).

FIGURES

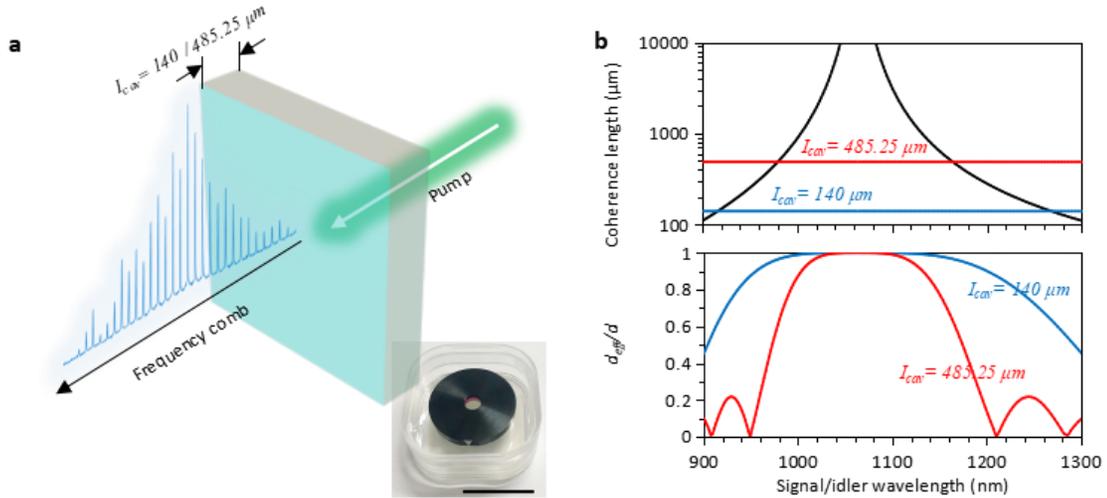

**FIG. 1 The schematic of the SOPO comb generation. a.** The SOPOs for the comb generation. Two types of SOPOs are fabricated with cavity lengths of 140 *μm* and 485.25 *μm*, respectively. Inset shows the picture of our SOPO sample with aluminum mount. **b.** The coherence length and $d_{eff}$ calculation. In the upper figure, the solid blue and red lines mark the CPM condition with $l_{Cav}=l_{Coh}$ for different cavity lengths. In the lower figure, blue and red lines shows $d_{eff}$ as a function of signal/idler wavelengths for $l_{Cav}=140$ *μm* and 485.25 *μm*.

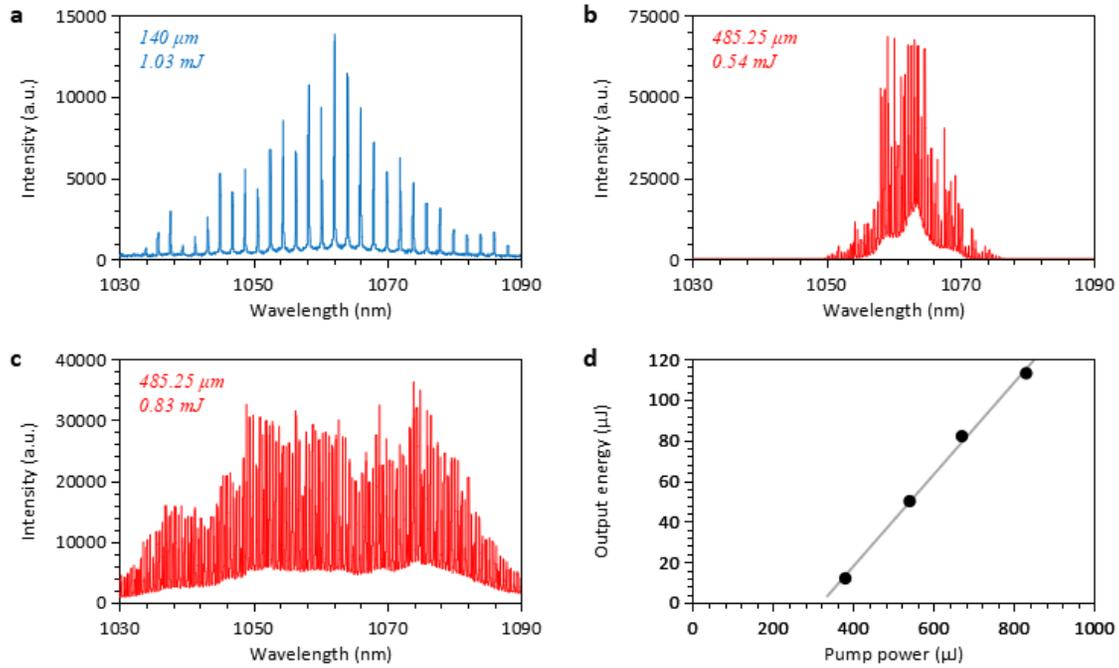

**FIG. 2 Comb spectrum and output pulse energy from SOPOs.** The spectrum of **a.** 140 *μm* SOPO at 1.03 *mJ* pump. **b.** 485.25 *μm* SOPO at 540 *μJ* pump and **c.** at 830 *μJ* pump. **d.** The output pulse energy as a function of pump pulse energy. 130 *μJ* output power is measured with a high slope efficiency of 22.6 %.

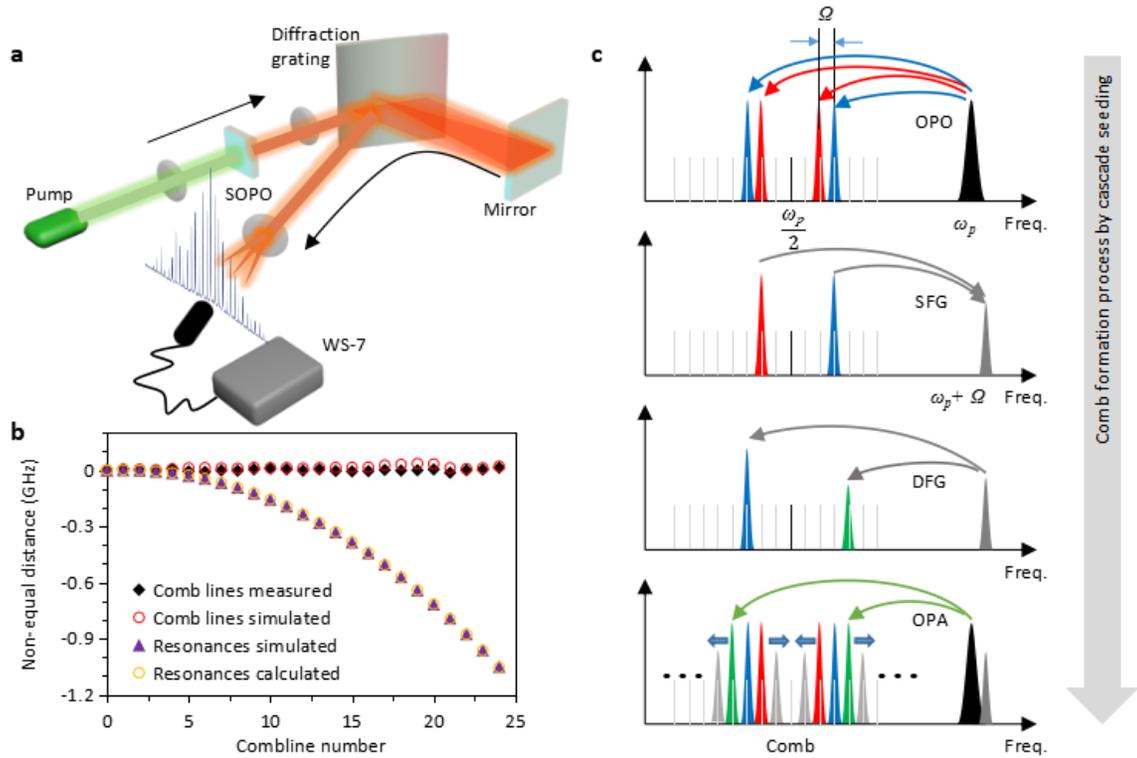

**FIG. 3 Line-to-line measurement of the SOPO comb. a.** Experiment setup to measure the optical frequencies of individual comb lines. **b.** Comb line and μOPO resonance frequency deviations from equal spacing. Error bars shows the instrument limit of wavelength meter. The comb lines #22-24 are in high order transverse mode and can only be detected via multi-mode fiber, which results in a reduced accuracy. **c.** Schematic of the cascaded $\chi^{(2)}$ line-to-line interaction. The red, blue and green pairs denote the comb lines #+/-i, #+/-(i+1), #+/-(i+2). Stating form comb line pairs #+/-i and #+/-(i+1), comb line pair #+/-(i+2) can be generated via cascaded SFG, DFG and OPA processes. Such processes can be repeated for a coherent comb generation over the whole span.

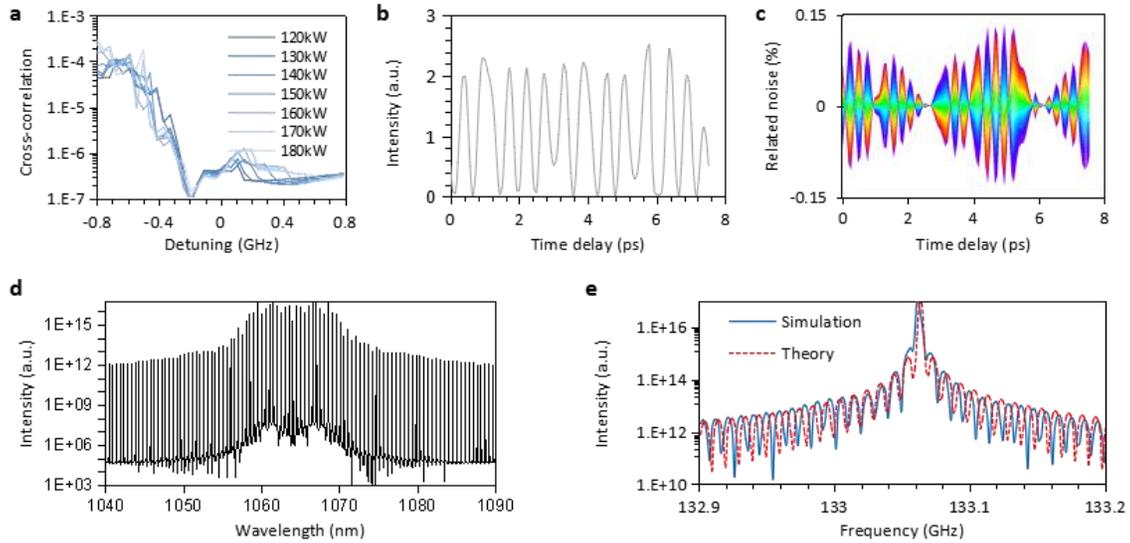

**FIG. 4 The simulation of the comb generation under CW pump. a.** The noise level as a function of pump detuning at different pump power. The lowest noise can be achieved at -193 *MHz* pump detuning and 12 *kW* pump, where the spectral and temporal behaviors are studied in detail. **b.** The temporal profiles of 266 repetitions. **c.** The relative intensity noise of 266 repetitions. **d.** The output spectrum of continue wave pump. **e.** The beat-note spectrum of the RF peak around the repetition frequency.